\documentclass[letter]{aa} 
\usepackage{graphicx}
\usepackage{float}
\usepackage{txfonts}
\usepackage{amssymb}
\usepackage{natbib}
\usepackage{url}
\defcitealias{D2023}{D2023}
\defcitealias{WHAM}{WHAM}
\defcitealias{VTSS}{VTSS}
\usepackage{amstext}

\begin{document}

   \title{Velocity measurement in the extensive [OIII] emission region 1.2$^{\circ}$ south-east of M31\thanks{Observations done at Haute-Provence Observatory}}
\authorrunning{Amram et al.}
\titlerunning{velocity in the extensive [OIII] emission region 1.2$^{\circ}$ SE of M31.}

\author{
P. Amram \inst{1}
\and
C. Adami \inst{1}
\and
B. Epinat \inst{1,2} 
\and
L. Chemin \inst{3}
          }

   \institute
        {Aix-Marseille Univ., CNRS, CNES, LAM, 38 rue Frédéric Joliot Curie, 13338 France\\
              \email{philippe.amram@lam.fr}
         \and
         Canada-France-Hawaii Telescope, 65-1238 Mamalahoa Highway, Kamuela, HI 96743, USA
         \and
         Instituto de Astrofisica, Universidad Andres Bello, Fernandez Concha 700, Las Condes, Santiago RM, Chile
             }
   \date{Accepted: March 7,  2023}

 
  \abstract
{The discovery of a broad, $\sim$1.5$^{\circ}$ long filamentary [OIII] 5007 emission $\sim$1.2$^{\circ}$ south-east of the M31 nucleus has recently been reported. More than 100 hours of exposures of a wide field (3.48$^{\circ} \times 2.32^{\circ}$) have allowed this pioneering detection based on 30 \AA\ narrow-band filters and several small refractors equipped with large cameras.}
 {We report a first velocity measurement in this extensive [OIII] emission line region.} 
   {We used the low-resolution spectrograph MISTRAL (R $\sim$ 750), a facility of the Haute-Provence Observatory 193 cm telescope. The velocity measurement is based on the H$\alpha$, [NII], [SII] and [OIII] lines.}  
   {The best solution to fit the spectrum indicates that the H$\alpha$ and [OIII] emissions are at the same heliocentric line-of-sight velocity of -96$\pm$4 km s$^{-1}$. This was measured within an area of $\sim$250 arcsec$^2$ selected on a bright knot along the long filament of $\sim$1.5$^{\circ}$, together with a [OIII]5007 surface brightness of 4.2$\pm$2.1 10$^{-17}$ erg s$^{-1}$ cm$^{-2}$  arcsec$^{-2}$. This agrees moderately well with the previous measurement. We also estimated the  H$\alpha$/[NII] line ratio as $\sim$1.1.}
   {The radial velocities at which the H$\alpha$ and [OIII] lines were detected seem to show that these hydrogen and oxygen atoms belong to the same layer, but we cannot exclude that another weaker [OIII] line, belonging to another structure, that is, at another velocity, is below our detection threshold.
   Different scenarios have been considered to explain this filamentary structure.  The extra-galactic origin was excluded in favour of Galactic origins.  
  We tentatively assume that this filament is a piece of a supernova remnant located at a distance of $\sim$0.7 kpc from the Sun, of which we only see a small fraction of the shells with a radius of $\sim$35 pc. The progenitor may be along the line of sight of the galaxy M31, but this observation might also just be part of a large-scale filamentary structure that should be investigated further.}


   \maketitle
%

\section{Introduction}

Filter widths at half-maximum ranging from $\sim$940 \AA\ (B/b-band) to $\sim$1480 \AA\ (R/r-band) are used in most optical wide-field sky surveys. This allows detecting stellar continuum emission in stars or in galaxies, but not the intrinsic narrow emission line, as was first done by the Sloan Digital Sky Survey \citep[SDSS;][]{York_2000} and by many others since.  
Medium-band filters are used in multiple narrow-band cosmological surveys that are carried out using many filters. 
They are typically ten times narrower than broad-band surveys, such as the Calar Alto Deep Imaging Survey  \citep[CADIS;][]{Wolf_2001a, Wolf_2001b}, 
and since, many others as well like the Javalambre-Physics of the Accelerated Universe Astrophysical Survey \citep[J-PAS;][]{Benitez_2014}.
J-PAS uses  54 filters with a width of  145 \AA, placed  100 \AA\  apart over a multi-degree field of view (FoV).
Medium-band filter surveys are more efficient in measuring the photometric redshift of distant galaxies than extended emission-line regions.
Indeed, even a medium-band filter is not yet narrow enough to detect the Balmer lines emitted by the hydrogen atom or auroral lines emitted by several atomic species (O, O$^+$, O$^{++}$, N$^+$, and S$^{++}$), which are drowned in the diffuse sky background.
Narrow-band surveys (FWHM = 10-20 \AA) that allow detecting emission lines are much more time-consuming because almost each emission line needs its proper narrow band at each redshift. To follow the same line when it is redshifted, a wide spectral range typically needs to be followed in steps of 10-15 \AA. For example, to detect a Balmer line or an auroral line from z=0 to z=0.1 with a constant velocity step (i.e. a constant resolution), we need $\sim$40 narrow-band filters per emission line with an average step of 15 \AA, increasing linearly with the wavelength. In addition, 
medium-band filters should be coupled to broad-band imagery to remove the continuum emission.  For this reason, wide-field emission-line surveys are rare and are therefore often conducted at low spatial resolution to reduce the observing time.
Two narrow-band filter surveys almost cover the whole sky :
(1) the Virginia Tech Spectral line Survey \citep[VTSS;][]{Finkbeiner_2003} covered the northern hemisphere with a narrow bandpass (17 \AA) H$\alpha$ filter at a resolution of 6' and a usable radius of 5$^{\circ}$ for each pointing and (2) 
the Southern HAlpha Sky Survey \citep[SHASSA;][]{Gaustad_2001} mapped a latitude south of 20$^{\circ}$, with 13$^{\circ}\times$13$^{\circ}$ FoVs and a resolution of 1.6' and 4.0', depending on the sensitivity. 
Alternatively, the Wisconsin H-Alpha Mapper  Northern Sky Survey \citep[WHAM;][]{Reynolds_2002, Reynolds_1998, Haffner_1998, Haffner_1999, Haffner_2003} covers the sky north of declination -30$^{\circ}$ with an angular resolution of $\sim$1$^{\circ}$ and a sensitivity of 0.15 Rayleigh. It consists of 37,565 spectra obtained with a dual-etalon Fabry-Perot filter instead of narrow-band filters.  The velocity resolution of $\sim$12 km s$^{-1}$ over a velocity range of $\sim$-90, +90 km s$^{-1}$ enables the removal of the geocoronal H$\alpha$ emission.  The  \citetalias{WHAM} survey has shown that interstellar  H$\alpha$ emission is detected in the whole sky, with intensities that range from thousands of Rayleigh near the Orion nebula and hundreds of Rayleigh in a large HII region (e.g. Barnard's loop) to 0.5 Rayleigh in faint high-latitude regions.  

\citet[][hereafter refereed to as \citetalias{D2023}]{Drechsler_2023} used a [OIII] 5007 \AA\ narrow-band filter (FWHM = 30 \AA) on a 106 mm refractor and accumulated wide-field exposures of M31 during a total of 24.6 hours at various dark observing sites in Lorraine, France. Confirming observations were obtained using a 106 mm and a 135 mm telescope in California and in New Mexico, cumulating an additional 85.5 hours and 24.9 hours in [OIII], respectively. This team of astronomers reported the discovery of an unknown broad,
1.5$^{\circ}$ long filamentary emission nebulosity 1.2$^{\circ}$ south-east
of the M31 nucleus based on the large FoV of 3.48$^{\circ} \times 2.32^{\circ}$ that is allowed by small telescopes and large cameras. 
On their website\footnote{\url{https://www.astrobin.com/1d8ivk/}}, these authors show [OIII] and H$\alpha$ images around M31 that show that the H$\alpha$ and [OIII] shapes of the flux distribution are very different. Along the same line of sight (LoS) lies H$\alpha$ emission that follows the large-scale patchy distribution, and also a filamentary and linear [OIII] emission that resembles cirrus fibratus radiatus in meteorology, that is, displays a very narrow band of fibrous filaments. In addition, this [OIII] emission has no obvious emission counterparts from radio to X-rays wavelengths.
These authors estimated an [OIII] 5007 surface brightness of 4$\pm$2 10$^{-18}$ erg cm$^{-2}$  s$^{-1}$ arcsec$^{-2}$.  

In this paper, we report a velocity measurement in a region belonging to the filament that we compared to a region located outside the filament. Sect. \ref{Observations and data reductions} and Appendix \ref{data reduction} describe the observations and the data reduction.  Sect. \ref{Results} provides the results, which are discussed in Sect. \ref{Discussion} before we conclude in Sect. \ref{Conclusions}. 
  
\begin{figure}[ht]
\centering \includegraphics[width=0.48\textwidth]{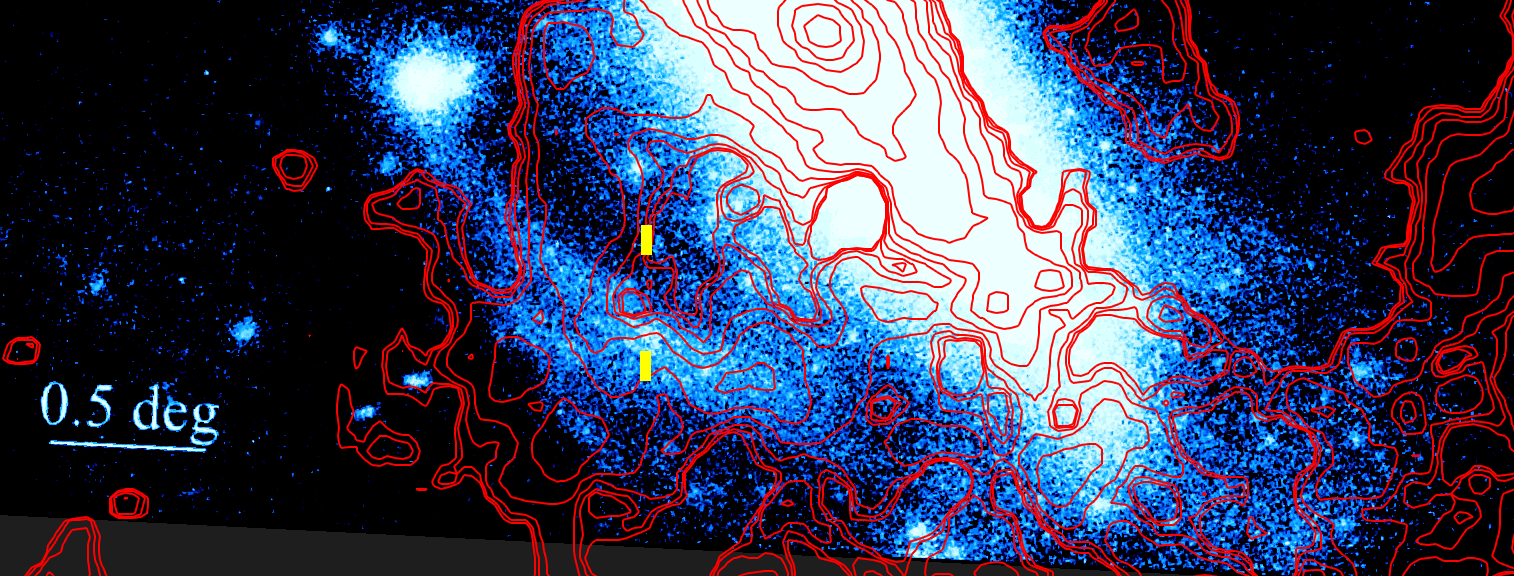}
\caption{Extended [OIII] 5007 emission region south-east of M31 from \citetalias{D2023}. H$\alpha$ isocontours from \citet[VTSS,][]{Finkbeiner_2003} are over-plotted in log scale (in red). These contours are described in Fig. \ref{OIII_VTSS_large}. The two yellow rectangles indicate the  MISTRAL long-slit positions, the one called "onset" is located to the south, and the comparison rectangle, called the "offset" slit, lies to the north, where both H$\alpha$ and [OIII] emissions are fainter.  Their widths are enlarged by a factor 10 for visibility.  North is up and east left.  The displayed FoV is $\sim$5.1$^{\circ}\times$1.6$^{\circ}$.}  
\label{OIII_VTSS_small}
\end{figure} 
  
\section{Observations and data reductions}
\label{Observations and data reductions}

Under excellent conditions of transparency and with a seeing of $\sim$2 arcsec, we have obtained long-slit spectra on January 25, 2023, using the low spectral resolution R = 700 at H$\beta$ and R = 750 at H$\beta$, which gives a line spread function (LSF); and a velocity dispersion of 2.95 \AA\ (182 km s$^{-1}$) and of 3.72 \AA\ (170 km s$^{-1}$) at H$\beta$ and H$\alpha$ wavelengths, respectively. The MISTRAL\footnote{Multi-purpose InStrument for asTRonomy At Low resolution} spectrograph\footnote{\url{http://www.obs-hp.fr/guide/mistral/MISTRAL_spectrograph_camera.shtml}} installed on the 193 cm OHP telescope is equipped with a blue grism that allows covering the wavelength domain ranging from 4250 to 8000 \AA\  \citep{Adami_2018}. 
We spent one hour on the [OIII] filament detected by \citetalias{D2023}, hereafter referred to as the onset spectrum, and one hour offset from it, referred to as the offset spectrum.  The 1D science spectra around the lines of interest are shown in Figs. \ref{finalspectrum} and  \ref{OIII}. Details about the data reduction are given in Appendix \ref{data reduction}.

\section{Results}
\label{Results}

\begin{figure*}[ht]
\centering \includegraphics[width=0.49\textwidth]{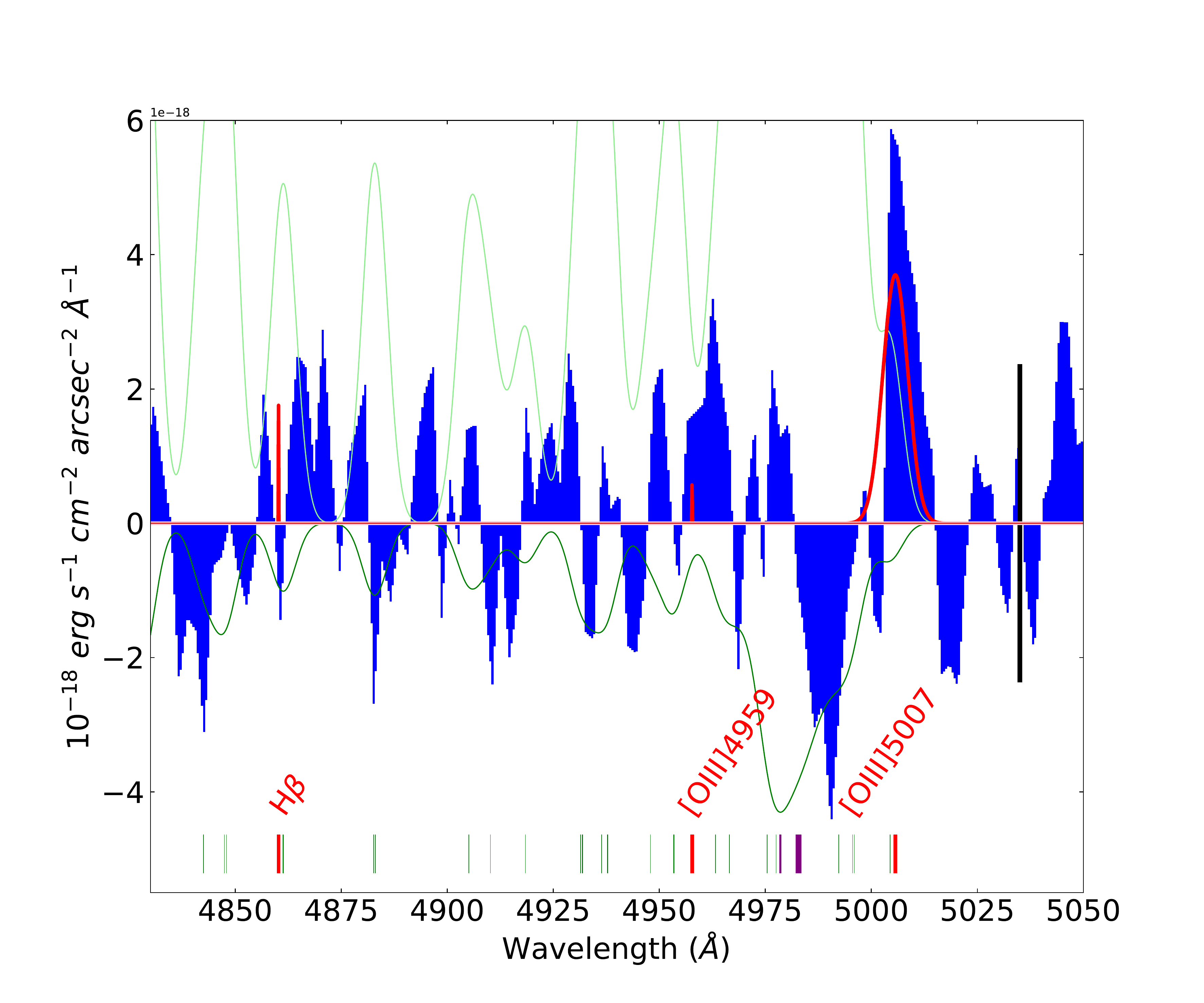}
\centering \includegraphics[width=0.49\textwidth]{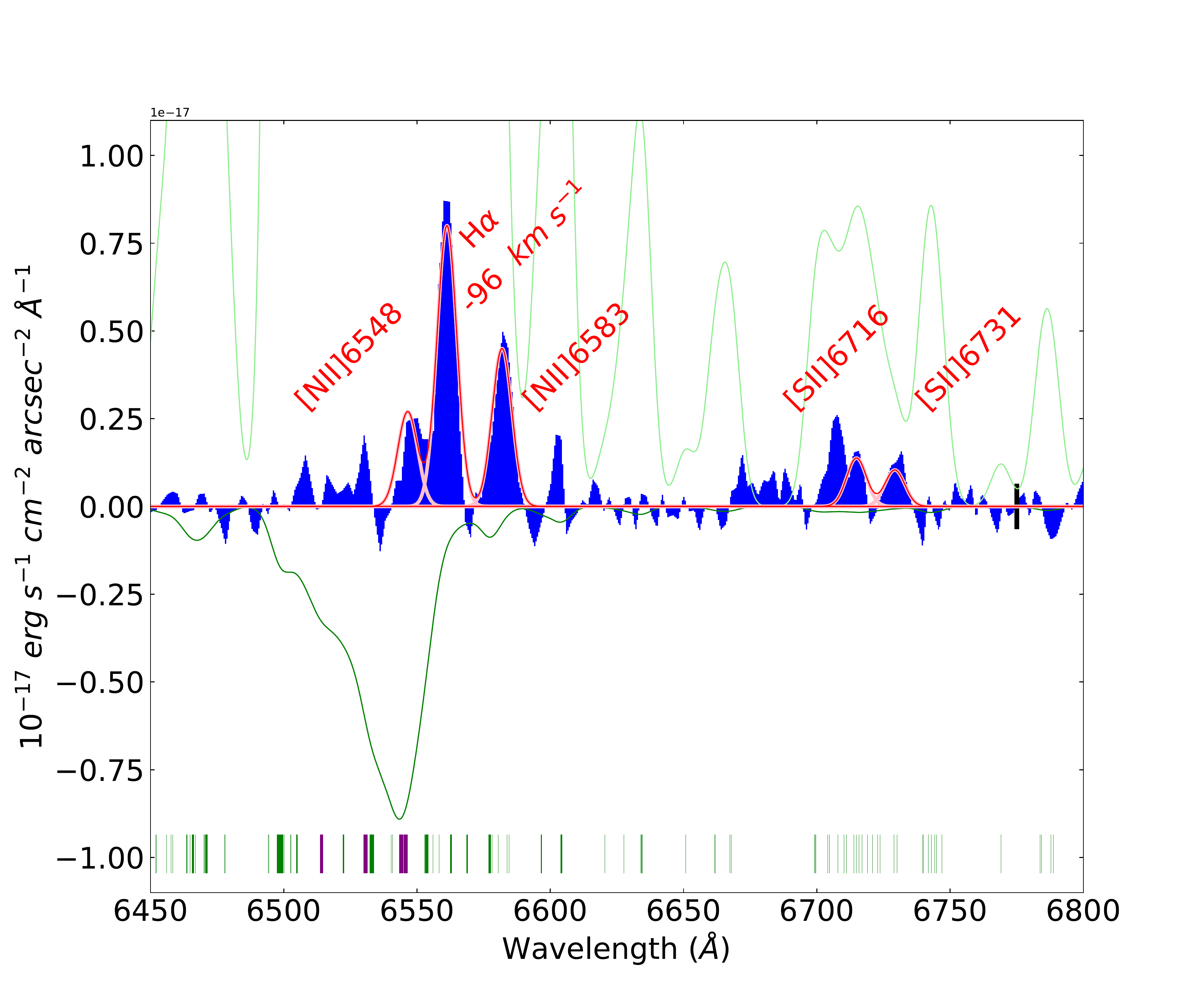}
\caption{
Wavelength and flux calibrated science spectra (blue histograms) in the blue (left panel) and red (right panel) bands.  The expected locations of the blueshifted H$\beta$, [OIII], H$\alpha$, [NII], and [SII] lines have been fitted by Gaussian function (pink lines). The red line is the sum of the Gaussian functions. The small vertical bars below the spectra indicate the night skylines from UVES (in green) and sodium emission lines due to urban light pollution of human origin (in purple), their widths illustrate their relative intensity. The sum of all those lines is over-plotted in light green.  Because the night skylines extend beyond the frame, they have been downscaled by a factor 5 and 50 for the blue and the red spectrum, respectively, and are mirrored symmetrically to the x-axis to fit in the frame (in green). The bold vertical black error bar on the right side of the plots indicates the standard deviation of the continuum between the emission lines.
}
\label{finalspectrum}
\end{figure*}

When all the identifiable lines are fit together to optimise the measurement, the best solution gives a heliocentric LoS velocity of -96$\pm$4 km s$^{-1}$, meaning that all the chemical elements belong to the same layer.
The fluxes in H$\alpha$ and [OIII] 5007 are 7.5$\pm$2.5 and 2.7$\pm$1.4 10$^{-17}$ erg s$^{-1}$ cm$^{-2}$  arcsec$^{-2}$  (or 13.2 and 3.7 Rayleigh), respectively. 
Our [OIII] flux measurement is about seven times higher than that of 4$\pm$2  10$^{-18}$ erg s$^{-1}$ cm$^{-2}$  arcsec$^{-2}$ from \citetalias{D2023}.  Considering that we chose to observe a bright knot both in the [OIII] and H$\alpha$ maps, our [OIII] flux measurement agrees acceptably well with that of \citetalias{D2023}. The [OIII] surface brightness of the offset spectrum is 5.1$\pm$2.6  10$^{-18}$ erg s$^{-1}$ cm$^{-2}$  arcsec$^{-2}$, which is less than twice larger than the \citetalias{D2023} mean measurement. This  suggests that the [OIII] filament is probably somewhat more extended than detected in narrow-band imaging.
The H$\alpha$/[NII]6548, H$\alpha$/[NII]6583, and H$\alpha$/[NII]6548+6583 line ratios are equal to 3.1, 1.7, and 1.1, respectively.  The LSF correctly fits all the spectral lines, which means that the lines are not resolved. The intrinsic dispersion of the emission lines is therefore lower  than 137-147 km\ s$^{-1}$.   The H$\alpha$/H$\beta$ ratio is $\ge$ 4.6.

\begin{figure}[ht]
\centering \includegraphics[width=0.48\textwidth]{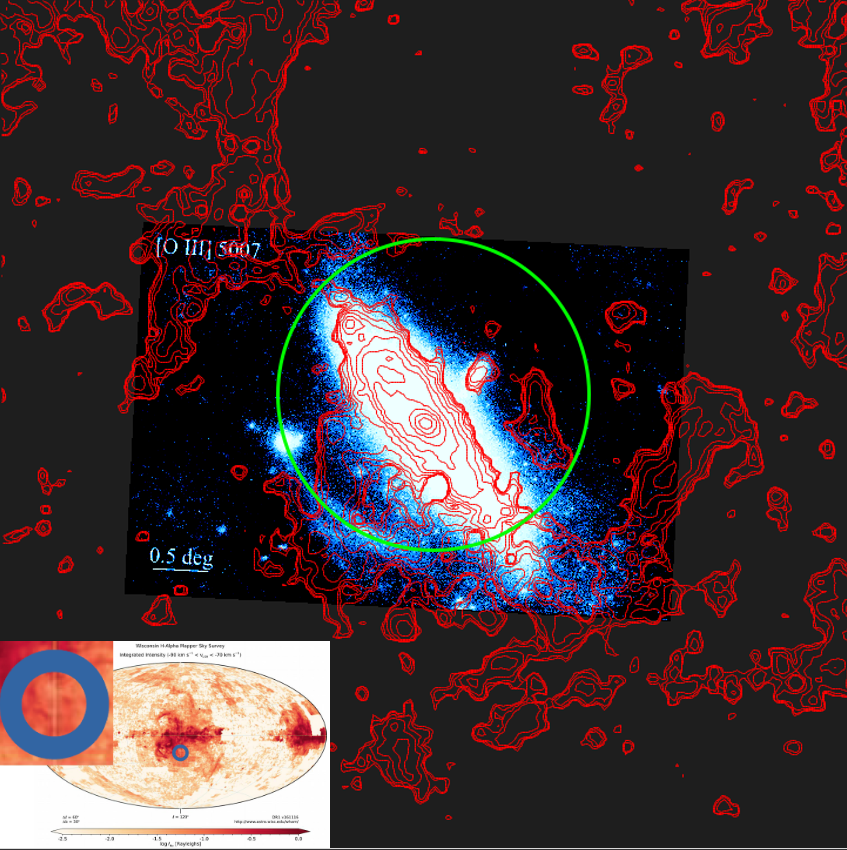}
\caption{Same as Fig. \ref{OIII_VTSS_small}, but on a wider FoV of $\sim$8.0$^{\circ}\times$8.0$^{\circ}$, showing the large-scale  \citetalias{VTSS} H$\alpha$ filaments calibrated with  \citetalias{WHAM}. The isocontours are given in log scale. They have the following increasing intensities in Rayleigh: 2.60, 2.65, 2.74, 2.92, 3.25, 3.87, 5.03, 7.20, 11.27, 18.89, 33.18, and 59.95. The green circle extrapolates the shape of the [OIII] filament. The centre of the circle indicates the possible location of a foreground central stellar progenitor along the LoS of M31.  The bottom left inset shows the  \citetalias{WHAM}-integrated H$\alpha$ intensity, which ranges between -90 and -70 km s$^{-1}$.  The location of M31 at a Galactic latitude of $\sim$-22$^{\circ}$ is indicated by the blue circle, which in turn is zoomed in the upper left corner.}
\label{OIII_VTSS_large}
\end{figure}

\section{Discussion}
\label{Discussion} 

The first question that should be discussed is why this extremely extended structure was discovered only now. The answer was given by the discoverers and is described in the introduction: it is very weak, and thus requires very long exposures in narrow-band imagery and goes unnoticed when wider filters are used. This emission is unresolved and diffuse, thus the use of a telescope with a larger mirror would not change the detection limit. This discovery demonstrates the importance of using a modest collecting surface, which makes it possible to cover a very large FoV with large modern detectors. The second question is why this structure has been detected in [OIII] and not in H$\alpha.$  The answer is probably linked to the presence of night skylines  \citep[geocoronal H$\alpha$ plus OH lines: 12 lines ranging between -150 and + 120 km s$^{-1}$ around H$\alpha$ at rest;][]{Haffner_2003}, which are more intense around H$\alpha$ than around [OIII] 5007,  as well as the sodium lines due to possible urban light pollution.
All the spectral lines of interest (hydrogen, oxygen, nitrogen, and sulfur) are highly contaminated by the night skylines. Despite this, low-resolution spectrography makes it possible to measure some of them better than what could be done in imaging.  When we consider a H$\alpha$ flux of $\sim$13 Rayleigh, which is rather high, we estimated that the night skylines passing through an FWHM$\sim$30 \AA\ filter centred around H$\alpha$ at rest (which provides an equivalent power of resolution R$\sim$219) are more than 12 times more intense than the H$\alpha$ flux of interest.
The third question is the nature of the object. It might be an extragalactic or a Galactic structure. We discuss the first hypothesis, which seems to us the most improbable, in Appendix \ref{An extra-Galactic object ?}.

\subsection{Nearby supernova remnant?}

The strong [NII] lines observed here as in other SNRs might be due to nitrogen-enriched gas thrown off by the pre-supernova star. 
The [NII]6583/[NII]6548 line ratio is $\sim$ 1.8, which is lower than the canonical value of 2.8, but [NII]6548 is blended with H$\alpha$.  The uncertain [SII]6716/[NII]6731 ratio is equal to $\sim$1.3. The H$\alpha$/[OIII]5007 ratio is $\sim$ 1.5.
\citet{VandenBergh_1973} published a photographic atlas of 24 galactic supernova remnants (SNRs), 9 of which extend on more than 1$^{\circ}$. Most of them display somewhat complex spherical shapes that are embedded in each other (the most famous is S147, the crab, vela, and cygnus loop ), but some others exhibit linear filaments of various lengths (e.g.  HB 9 or VRO 420501). The faint HB 3 SNR displays only one linear filament that is so short that the size of the SNR was not evaluated. Other SNRs are very shallow and diffuse (e.g. KES 45).  To distinguish SNRs from HII regions and to measure their physical properties, spectra are needed.   \citet{Daltabuit_1976} and \citet{Dodorico_1977} led spectroscopic studies of
SNRs that were extracted from the atlas of \citet{VandenBergh_1973}, and we can compare them to our target.
The low values of the H$\alpha$/[NII] ratio of 1.1 and of the [SII]6716/[SII]6731 ratio of 1.3 indicate that it might belong to an SNR and not to HII regions.  The H$\alpha$/[NII] line ratios of the reference SNR sample indeed range between 0.40 and 2.18, with a median value of 1.29, and the [SII]6716/[SII]6731 line ratio ranges between 0.50 and 1.44, with a median value of 1.23. Both median values are very similar to our measurements.  \citet{Daltabuit_1976} showed different correlations between physical quantities (line ratios, diameters, and expansion velocities) that should be understood as the result of shock waves propagating in the interstellar medium. Firstly, the sulphur intensity ratio  [SII]6716/[SII]6731 and H$\alpha$/[NII] are correlated. 
Our measurements follow the trend given by this correlation. Secondly, two correlations exist between the intensity line ratios and the diameter of the SNR,  which allows us to guess one diameter measurement and thus a distance estimate for our target.  The first correlation is between [SII]6716/[SII]6731 and the diameter, the second correlation is between H$\alpha$/[NII] and the diameter.  When we use these correlations as template values, the sulphur and H$\alpha$/[NII] line ratios would give a diameter of 45$\pm$5 pc and 35$\pm$5 pc, respectively. We would favour the last value, taking into account that the line ratio determination of H$\alpha$/[NII] is more robust that that of sulphur. The two measurements are nevertheless compatible.  When we tentatively extrapolate the shape of the filaments that are slightly curved, to guess a possible location for the SN progenitor (see Fig. \ref{OIII_VTSS_large}), the angular radius of the SNR would be $\sim$1.4$\pm0.1^{\circ}$. This leads to a distance of the target of 0.7$\pm$0.1 kpc from the Sun.
Thirdly, a correlation exists between the H$\alpha$/[NII] line ratio and the expansion velocity of the SNR.    
The expected H$\alpha$/[NII] line ratio for an expansion velocity of the SNR of 96 km\ s$^{-1}$ is $\sim$1.2$\pm$0.1, which is fully compatible with our measurement.  
Finally, we may wonder where the progenitor of the SNR is. It is probably a low-magnitude evolved star, and if the drawing that sketches the shape of the SNR shell is to believed, it might be undetectable if it is along the LoS of M31.
 
\subsection{Galactic filament?}
\label{A Galactic filament}

Fig. \ref{OIII_VTSS_large} displays the large-scale  \citetalias{VTSS} H$\alpha$ filaments in a larger FoV of $\sim$8.0$^{\circ}\times$8.0$^{\circ}$, surrounding the $\sim$3.5$^{\circ}\times$5.1$^{\circ}$ \citetalias{D2023} [OIII] image.  Fig. \ref{OIII_VTSS_small} shows that H$\alpha$ emission is also present south-east of M31, where [OIII] emission has been detected.  This is confirmed by Fig. \ref{OIII_VTSS_large}, which shows that the H$\alpha$ filament that matches the [OIII] emission is extended south-west of M31,  to trace almost half a ring, similar to Bernard's loop (Sh 2-276). This figure also exhibits other filaments at larger distances from M31, which expand over much larger FoVs than displayed in Fig. \ref{OIII_VTSS_large}. Even though the [OIII] emission has been detected on scales of several degrees, H$\alpha$ emission has previously been detected on much larger scales.   If the [OIII] and H$\alpha$ emissions are in some way correlated,  it is difficult to interpret the [OIII] emission without knowing whether [OIII] continues to follow the H$\alpha$ emission outside the  \citetalias{D2023}  FoV.

The  \citetalias{WHAM} and  \citetalias{VTSS} surveys we described in the introduction have shown that interstellar  H$\alpha$ emission is detected in the whole sky, with intensities that range from thousands of Rayleigh near the Orion nebula and hundreds of Rayleigh in large HII regions (e.g. Barnard's loop) to 0.5 Rayleigh in faint high-latitude regions.  Interstellar H$\alpha$ emission fills the sky with loops, filaments of any sizes, and unresolved source ($<$1$^{\circ}$) or large emission enhancements, including large-scale filaments that are superposed on a diffuse H$\alpha$ background that is essentially due to the Galactic free-free emission \citep[e.g.][]{Marcelin_1998} in addition to the complex geocoronal emissions \citep{Nossal_2001}.  
 \citetalias{WHAM} is a spectrometric survey that provides a flux and wavelength profile for each squared degree, from which a  night skyline-free barycentric velocity is computed.  The H$\alpha$ velocity distribution of the northern sky is dispatched in lambda maps in step of 20 km s$^{-1}$.  The maximum intensity of the squared-degree region that includes the \citetalias{D2023} [OIII] image is found in the lambda map [-90,-70] km s$^{-1}$ (shown in the small inset in the bottom left corner of Fig. \ref{OIII_VTSS_large}, where the location of M31 is indicated by a blue circle). We know that  \citetalias{WHAM} does not extend at velocities lower than -90 km s$^{-1}$, and therefore, our measurement of -96$\pm$4 km s$^{-1}$ agrees very well with it.  At this scale, the H$\alpha$ flux encircled in this blue circle  looks homogenous. Using the  \citetalias{VTSS} data within $\sim$1$^{o2}$ around the target, we estimated the mean H$\alpha$ flux and its standard deviation to 3.8$\pm$0.3 Rayleigh.  The H$\alpha$ flux measured from Fig. \ref{finalspectrum}, which is the difference between the onset and the offset regions, is  $\sim$14.5 Rayleigh.  When we hypothesise that the flux of the offset region is given by the mean flux measured from  \citetalias{VTSS} and calibrated using  \citetalias{WHAM} (i.e. 3.5 Rayleigh), our flux measurement is $\sim$3.8 times higher than the diffuse  \citetalias{WHAM} H$\alpha$ emission. We chose to observe a knot in the [OIII] image, however, which may also correspond to a knot in H$\alpha$ . \citetalias{WHAM} and  \citetalias{VTSS}, with their resolutions of 1$^{\circ2}$ and 6'$^2$ , respectively, cannot resolve the tiny FoV covered by our slit, which is $\lesssim$ 2 10$^{-5}$ and  300 times smaller, respectively. 
Many faint filaments 
have no obvious correspondence to any previously known structures or to the other phases of the interstellar medium revealed by 21 cm radio continuum surveys, IR, or X-ray observations,  and they have no readily identifiable origin or source of ionisation. Significant kinematical variations are also observed among various features. 
\citetalias{D2023} 
argued that the surface brightness distributions of H$\alpha$ and [OIII] are very different (see \ref{A Galactic filament}), and this is what we can observe from their images\footnote{\url{https://www.astrobin.com/1d8ivk/}}. An inspection  also show linear features in this H$\alpha$ image, however, that are much less extended and spectacular than the fine and long [OIII] filaments. Some have the same orientation as the [OIII] filaments, however.
A diffuse ionisation source is probably required to maintain constant intensities along the filaments.  Many mechanisms may produce large and faint filaments. The efficiency of leaking Lyman-continuum radiation from the disc is a long-standing problem for ionising a thick layer of the Galaxy and even the Galaxy halo  \citep{Bland-Hawthorn_1998}, but large-scale, kinematic Galactic structures such as Galactic rotation, superbubbles, chimneys, or worms  \citep{Koo_1992} may be at their origin as well.  The gas temperature (ranging from 6,000 to 10,000 K) and ionising states, which are requested to select the correct mechanisms, are rarely known because they require time-consuming additional line observations ([NII], [SII], and [OIII] lines).

 \section{Conclusions}
  \label{Conclusions}
 
The faint and extended almost linear filamentary region discovered by \citetalias{D2023} seems very unusual.  More precisely, extended [OIII] emission line regions over squared degrees like this have rarely been observed, even at redshift zero. This does not mean that they do not exist.  They have to be searched for in almost blind survey mode over a large-scale area to explore different physical mechanics and to lower the cosmic variance.    

The low [NII]/H$\alpha$ ratio of our data is most intriguing. It  allows us to speculate that this filament may be a piece of an SNR. However, the arguments to  support this speculation are not strong enough to confirm without ambiguity that this explains the origin of this emission, in particular the origin related to the curvature of the filament, which makes it possible to speculate on the size and distance of the SNR.  Further investigations are required to study different Galactic large-scale scenarios.

According to our velocity measurements, the structure detected in H$\alpha$ and in [OIII] belongs to the same layer. It is unclear whether the structure observed by \citetalias{D2023} is different from the detected structure. On the one hand, it is indeed possible that the line [OIII] that we have measured is detectable everywhere where H$\alpha$ has been seen in  \citetalias{VTSS} and  \citetalias{WHAM}, precisely because we have chosen to position the OFF slit in a region of weak emission in H$\alpha$ and in [OIII]. On the other hand, because the flux we measured is $\sim$6.8 times more intense than that detected by \citetalias{D2023}, it is possible that we were limited by the sensitivity of the instrument as well as by an insufficient exposure time.  Detecting and recognising a weaker structure in [OIII] at another LoS velocity is extremely challenging even when the two lines of [OIII] 4559 and 5007 are detected simultaneously. In any case, if this structure at another velocity exists, it is not detectable with our instrumental setup. We should aim to multiply the S/N by a factor of about five and to choose a resolving power $\text{about }$twice as high to overcome the very constraining lines of the night sky.

A seeing-limited, multiple narrow-band filter survey, including the two most important Balmer lines H$\alpha$ and H$\beta$ and auroral lines (oxygen, nitrogen, and sulphur) over a very wide field of $\sim$10$^{\circ}\times$10$^{\circ}$ centred around M31 would allow us to proceed in the understanding of the origin of this object, of the physical parameters (temperature and ionisation states) of the filaments that seem even more extended that what was observed by \citetalias{D2023}, when possible correlations between [OIII] and H$\alpha$ emission are searched for.  The H$\alpha$ observations from  \citetalias{VTSS} and  \citetalias{WHAM} around M31 (see Fig. \ref{OIII_VTSS_large}) show very extended structure that is more or less filamentary at the low  spatial resolution of WHAM. Unfortunately, in this FoV, neither  \citetalias{VTSS} nor  \citetalias{WHAM} explore other emission lines such as [OIII], and even if other lines were available,  \citetalias{WHAM} lacks a sufficient spatial resolution to resolve the filaments. 

There are magnificent and promising new opportunities for amateur astronomers armed with time and powerful wide-field observation facilities through small-size telescopes and their intrinsic large FoV, which are today equipped with large detectors at relatively affordable prices and a series of narrow filters whose prices have also been greatly reduced. Professional 3D surveys could also be conducted using fast Fourier transform interferometers or Fabry-Perot tunable filters, which would have the advantage of being able to cover the Balmer and auroral lines alone for a wide range of velocities and not the unneeded spectral ranges separating them. Their versatility in modulating the spectral resolution to separate the [NII] doublet from the H$\alpha$ line, for example, to resolve the [SII] doublet and so on, which are fundamental for all diagnostics, and in increasing the spectral resolution, radial and velocity dispersion velocities can be measured, which are also essential for understanding the observations. The future instrumental revolution will come from the use of multi-spectral detectors such as the Microwave Kinetic Inductance Detectors (MKIDs)\footnote{\url{https://web.physics.ucsb.edu/~bmazin/mkids.html}}, whose increasing power of resolution already today spans from R$\sim$35 at 2540 \AA\ to $\sim$14 at 13100 \AA. The growing size of their matrix, which will be mosaicable, will enable using all the orders of the tunable filter to cover the spectral range 2540 to 13100 \AA during a single scanning cycle of the interferometer. This will vastly increase the merit factor of the instrument.

\begin{acknowledgements}
The authors warmly thank Delphine Russeil for interesting discussions on the origin of filaments. Jérôme Schmitt is also kindly thanked for his assistance during the observations and for the setup of the MISTRAL instrument.  We are also grateful to the night operators for their assistance. We also thank Isabelle Boisse and Flavien Kiefer for letting us use 2 hours or their SOPHIE run, as well as the directors of the OSU (Jean-Luc Beuzit) and the OHP (Marc Ferrari) for the DDT allocation. This research is based on observations made with the T193/MISTRAL spectrograph and imager at Observatoire de Haute Provence (OHP, CNRS), France and has made use of the MISTRAL database, based at OHP, and operated at CeSAM (LAM), Marseille, France.
 \end{acknowledgements}

%
%

\bibliographystyle{aa} 
\bibliography{biblio} 

\begin{appendix} 

\section{Data reduction}
\label{data reduction}

\begin{figure*}
        \includegraphics[width=0.99\textwidth]{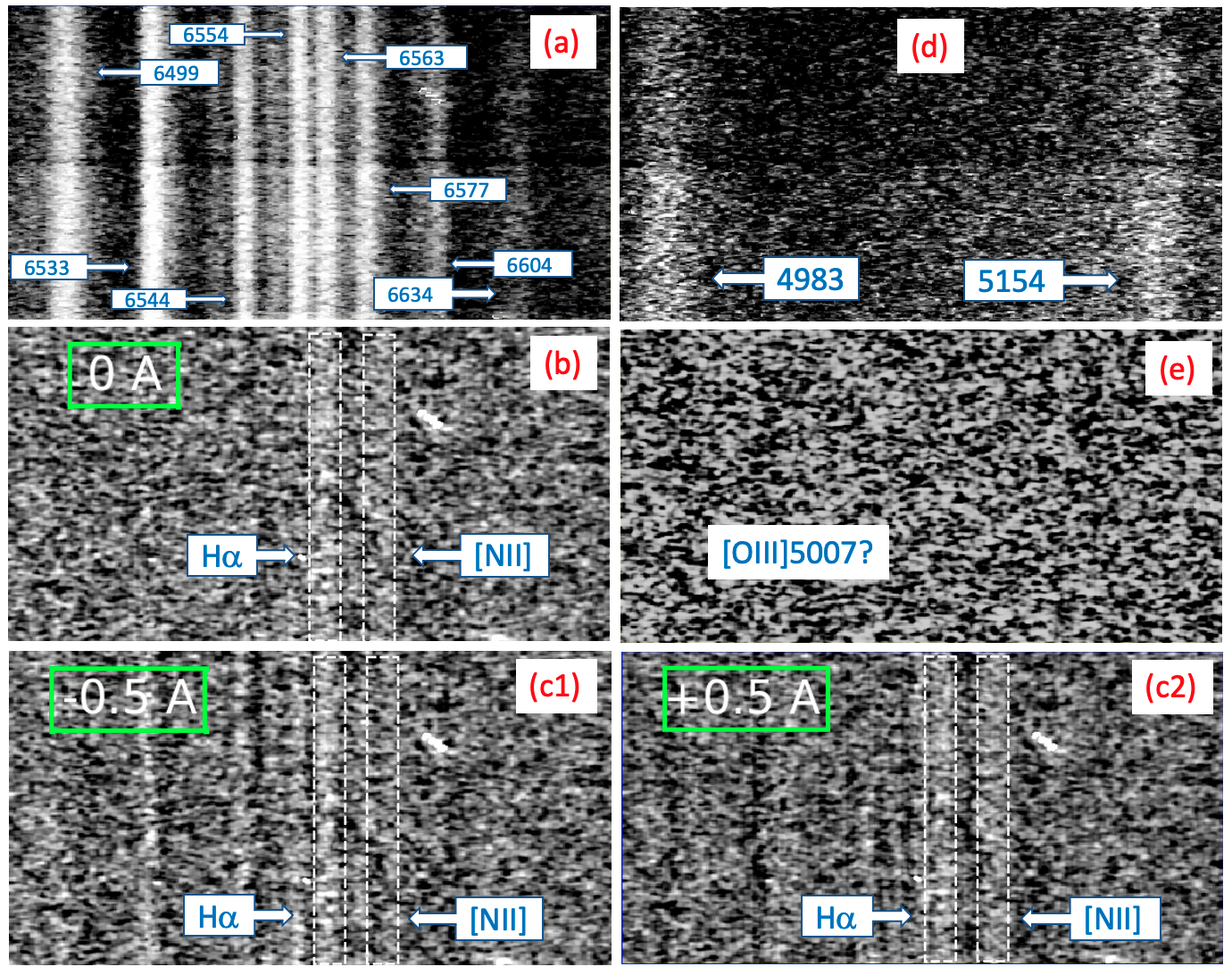}      
        \caption{2D spectra. Panels (a), (b), and (c): Red part of the spectrum around the H$\alpha$ and [NII] 6548 and 6583 lines. Panels (d) and (e): Blue part of the spectrum around [OIII] 5007. Panel (a): Raw spectrum containing the night skylines. Panel (b): Same as panel (a), but the night skylines are subtracted.  Panels (c1) and (c2): Position of the sky lines is shifted by -0.5 and + 0.5 \AA\  with respect to panel (b). The white dotted rectangle in panels (c1) and (c2) shows the position of the H$\alpha$ and [NII] 6583 lines as indicated by panel (b). Panel (d): Raw spectrum showing mainly two sodium night skylines. Panel (e): Same as panel (d). The night skylines are cleanly subtracted, but the [OIII] 5007 line is not visible in the 2D spectra.}
        \label{2D}
\end{figure*}

\begin{figure}
        \includegraphics[width=0.48\textwidth]{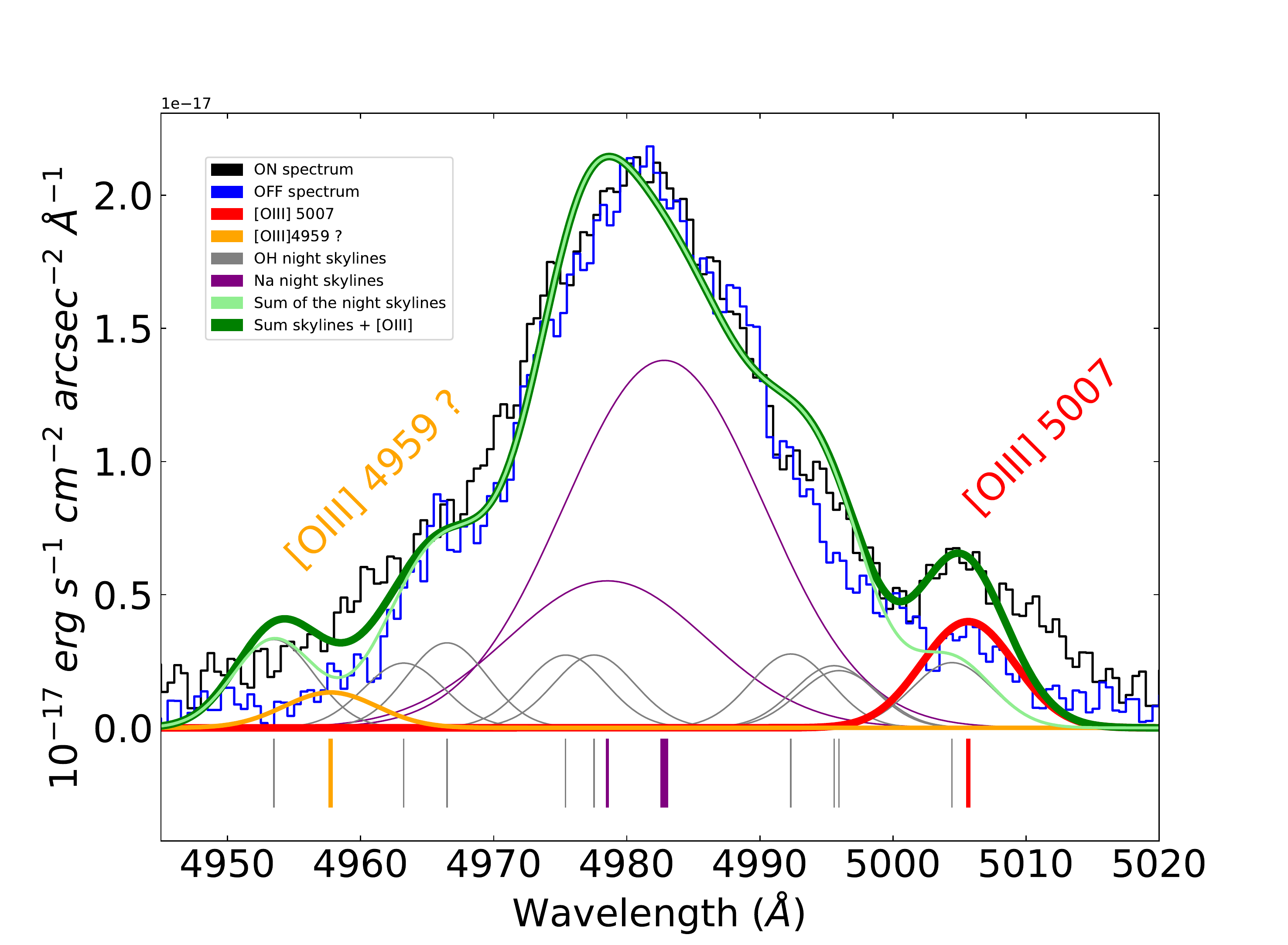}      
        \caption{1D spectrum around the [OIII]4959, 5007 lines. It is heavily polluted by urban lighting using sodium lamps.  Gaussian functions mimicking the UVES OH (in grey) and Na (in purple) night skylines are used to reproduce the sky emission of the offset spectrum (blue stairs).   The OH Gaussian line widths are identical and fixed by the LSF at [OIII]. The complex sodium line widths are identical, but free parameters of the fit. These latter line widths are equivalent to those observed by \url{https://astrojolo.com/wp-content/uploads/2020/01/night-sky-spectrum.png} at about the same resolution power.  The known wavelengths and the relative intensities of all the night sky lines are used to constrain the fit. The sum of all the night skylines is given by the  light green curve.  An [OIII] 5007 emission line (red line) is needed to fit the onset spectrum (black stairs), as displayed by the thick dark green curve, which is the sum of the thin light green and red curves. A Gaussian function with an amplitude of one-third of the Gaussian function of [OIII] 5007 is tentatively added to simulate the [OIII] 4959 line in order to fill the shift between the onset and offset spectra.  Similarly, an H$\beta$ line can used to fill the same type of broad shift at H$\beta$ between the on- and offset spectra (not shown).  The wavelengths of different lines are indicated by vertical lines
below the spectrum, using the same colour code.
}
        \label{OIII}
\end{figure}

We took four independent exposures, two located on the [OIII] filament, and two off the filament.  The observations were made following the pattern on-off-on-off. We waited 80 min after astronomical sunset for the intensity of the OH lines to settle close to its nocturnal value. The H$\alpha$/[NII] is the same, regardless of the pattern that excludes an instrumental or atmospheric effect. Raw and night skyline-subtracted 2D spectra around the H$\alpha$, [NII], and [OIII] lines are shown in Fig. \ref{2D}. Panel (b) shows that the [NII] 6583 is not superimposed on a night skyline.  It also convincingly shows the H$\alpha$ line, despite its weakness, which is not the case of the [NII] 6548 and  [OIII] 4959 and 5007 lines (panels b and e), which are not seen in the 2D spectra. The columns must have been collapsed to see them appear in the 1D spectra (see Figs. \ref{OIII} and \ref{finalspectrum}).  The sky residual in panels (c1) and (c2) illustrates that the best subtraction of the sky is indeed shown in panel (b), that is, the subtraction predicted by the calibration. Furthermore, we reduced the 2D spectra to 1D spectra by collapsing the information along the length of the slit using two different procedures: For the first procedure, the 2D on- and offset spectra were subtracted from each other before the difference was collapsed (see Fig.  \ref{finalspectrum}), whereas for the second procedure, the 2D spectra were collapsed into a 1D spectrum before they were compared (see Fig.  \ref{OIII}).
The useful length of the 1.9 arcsec wide slit was 224 arcsec, covering a FoV of 425 arcsec$^2$.  The positions of the two slits are shown (in yellow) in Fig. \ref{OIII_VTSS_small}, the location of the centre of the science slit is $\alpha(2000)$ = 0:45:49.53, $\delta(2000)$ = +40:09:13, and the location of the offset slit is $\alpha(2000)$ = 0:45:49.17, $\delta(2000)$ = +40:34:10.  In both cases, Galactic field stars were carefully avoided by pre-imaging, using the MISTRAL imaging mode and by prior SDSS r-band inspection of the FoV.  The data were wavelength calibrated using an arc calibration lamp, as described in \citet{Adami_2018}.  The wavelength calibration was checked using the night skylines measured by UVES \citep{Hanuschik_2003}. The background was subtracted using a spline function to flatten the spectra.

The 1D spectra around the lines of interest are displayed in Figs. \ref{finalspectrum} and  \ref{OIII}.  They were flux- and wavelength-calibrated using the procedures described in \citet{Adami_2018}. We used the star Hiltner600 as photometric standard. We locally estimated the continuum level and its noise $\sigma$, mainly due to night sky residuals, using the continuum spectrum by avoiding the emission lines.  We computed a standard deviation level of the noise  $\sigma \sim$ 0.64 and 2.37 10$^{-18}$ erg s$^{-1}$ cm$^{-2}$  arcsec$^{-2}$ \AA$^{-1}$ in the red and blue, as represented by the thick vertical black error bars in Fig. \ref{finalspectrum}. $\sigma$ is thus $\sim$ 3.7 times higher around H$\beta$ than around H$\alpha$.  
To better understand the intensity of the night skylines that exceed the limits of the plot, and in particular the intensity of sodium, we replotted these lines by inverting them with respect to the abscissa axis, and by decreasing their intensity by a factor of 5 and 50 in the blue and red parts of the spectrum, respectively, consistently with the sum of the two Na lines around [OIII] (i.e. 4979 and 4983), which are $\sim$40 times weaker than the sum of the four Na lines around H$\alpha$ (6514, 6531, 6544, and 6546). The total efficiency of the instrument is about four times lower at H$\beta$ than at H$\alpha,$ and in addition, we observe a strong increase in sky brightness in the blue in the spectra.  The combination of these two effects and the Balmer decrement H$\alpha$/H$\beta$ $\sim$  2.86, which might be even larger because of reddening, makes the detection of H$\beta$ and [OIII] 4959, 5007 more difficult than that of H$\alpha$ and [NII] 6548, 6583.

The expected fluxes are low and are therefore dominated by the sky background, which needs to be carefully subtracted. To do this, we chose an area outside the [OIII] and H$\alpha$ filament, but as close as possible to the science region. The strategy consisted of subtracting the offset spectrum from the onset spectrum with the aim of detecting a flux difference between the two regions. The fluxes detected in the science zone are therefore lower limits and not an absolute measurement.  We nevertheless tried to measure the surface brightnesses of the offset spectrum. This was possible for the [OIII] 5007 line, but not for the H$\alpha$ line due to the strong night skylines contamination.  Using Gaussian functions whose widths were fixed by the LSF (see Sect. \ref{Observations and data reductions}), we adjusted all the lines from the blue to the red part of the spectrum simultaneously with a single velocity by fixing the wavelength line separation at rest.  The global wavelength shift and the amplitudes of the lines were free parameters of the fit.
The solution presented in Fig. \ref{finalspectrum} corresponds to the best-fit model (lower $\chi^2$). The H$\alpha$ line and the [NII] 6548 and 6583 \AA\ doublet are clearly detected at $\sim$ 13.6, 4.3, and 7.9 $\sigma$ levels, respectively. It is more difficult to be convinced that the [SII] 6715 and 6727 \AA\ doublet is unambiguously detected because their wavelengths do not fit the expected locations perfectly because their S/N are at  the $\sim$1.6 and 2.1 $\sigma$ level. The numerous night skylines (plotted in grey) show how difficult the night skyline subtraction was.  In the bluer region of the spectrum (left panel of Fig. \ref{finalspectrum}), the [OIII] 5007 line is detected at $\sim$ 2.4 $\sigma$ levels. This line is not symmetrical because of the strong night skyline contamination at the blue edge of the line, which is too strong to allow recovering the integrity of the spectral lines, as can be seen from the strong dip in the blue wing. The H$\beta$ and [OIII] 4959 lines are below  $1\sigma$ level and are thus undetected. Their expected location on the spectrum is indicated by the pink Gaussian function. 
A heliocentric correction of -25 km\ s$^{-1}$ was added to the measured LoS velocity of -71$\pm$3 km\ s$^{-1}$ , giving a velocity of -96$\pm$4 km\ s$^{-1}$ (see Fig. \ref{finalspectrum}). 

\section{An extragalactic object?}
\label{An extra-Galactic object ?}

If these filaments belong to the intergalactic medium, given their angular proximity to Andromeda, the question is whether they are related to this galaxy. In this case, this gas may have been stripped during interaction with the Milky Way. The heliocentric velocity of M31 being -301$\pm$1  km\ s$^{-1}$ \citep{Watkins_2013}, if these filaments belong to M31, they have $\text{approximately}$ the size of the optical radius of M31 and are receding by $\sim$ 200  km\ s$^{-1}$ from its the systemic velocity. 
The Effelsberg-Bonn H I survey (EBHIS) comprises an all-sky survey north \citep{Kerp_2011}. HI emission is present everywhere around M31.  \citet{Kerp_2016} separated the extraplanar HI emission linked to M31 from the foreground Galactic emission. Their Fig. 4 shows that the brightest and most extended part of this extraplanar HI structure has an LSR velocity of $\sim$-118 km s$^{-1}$ towards the Galactic plane, that is, in the opposite direction of the [OIII] filament, where no HI is observed.
We only have one velocity measurement located in a random position along the filament. We therefore do not know whether this filament might present a velocity gradient that might be kinematically related in some way to the velocity gradient of M31  \citep[e.g.][]{Chemin_2009}.  
When we assume that the gas at the location of the filament is gravitationally bound to M31, its kinematics should thus roughly follow that of the M31 disc. By assuming a rotation velocity of $\sim 250$ km/s, which is the average rotation velocity of the HI gas in the M31 disc, at the location of the instrumental slit  (R $\sim 58$ kpc from the centre of M31 in the disc frame, and at an azimuthal angle of 95$^{\circ}$ with respect to the major axis, as measured using an inclination of 74$^{\circ}$, and a counter-clockwise position angle of the major axis of 38$^{\circ}$ from north), we estimate that the LoS velocity of the gas should be lower by -30 km s$^{-1}$  than the M31 systemic velocity at this location, that is, -330 km s$^{-1}$. This is significantly discrepant from the velocity measured for the ionised gas at the same position, however.
On the other hand, \citet{Madsen_2000} observed spatially extended H$\alpha$ distribution in M31 beyond the stellar disc.  They found an H$\alpha$ intensity of 0.12 Rayleigh beyond the bright stellar disc, but no significant detection within and off the outer HI disc. They placed an upper limit on the intensity of 0.04 Rayleigh, which is far lower than our detection. In addition, if this gas is stripped from M31, we would rather expect the filaments to be perpendicular to the plane of the M31 disc, whereas they are rather parallel and surprisingly straight and smooth and cover distances of several kiloparsec.
For the stellar component, the deep photometric survey of the Andromeda galaxy in the V- and i-band filters shows a panoramic view of the outermost region of the galaxy up to 150 kpc  \citep{Martin_2007}.  The inner giant stellar stream east, which is on the same LoS as the [OIII] emission, is  almost perpendicular to the [OIII] filaments. It should be helpful to try to detect [OIII] emission in the numerous stellar streams in the form of shells around M31 to study whether they are linked, which is unlikely.
In addition, if the stream comes from the interstellar medium of M31, no such low H$\alpha$/[NII] line ratio is expected. The gas content should essentially be made of hydrogen. It cannot be a supernova remnant (SNR) from a star belonging to M31 \citep[e.g. the first pulsar discovered in Andromeda;][]{Esposito_2016}  because the SNR would extend over more than 10 kpc, even 100 kpc for the complete structure of the shell, most of which would be invisible. At these large scales, the gas would be so diluted that it would not be detectable at the level at which it was detected. It is therefore probably necessary to seek a galactic origin for this diffuse filament, which is located -22$^{\circ}$ below the plane of the Milky Way (the Galactic latitude of M31).  It is also 33$^{\circ}$ above the ecliptic plane. This structure therefore does not belong to it.  

\end{appendix}

\end{document}